# Oxidation States in Solids from Data-Driven Paradigms

*Yue Yin, Hai Xiao**

Department of Chemistry, Tsinghua University, Beijing 100084, China

**ABSTRACT**

The oxidation state (OS) is an essential chemical concept that embodies chemical intuition but cannot be computed with well-defined physical laws. We establish a data-driven paradigm, with its implementation as Tsinghua Oxidation States in Solids (TOSS), to explicitly compute the OSs in crystal structures as the emergent properties from large-sized datasets based on Bayesian maximum a posteriori probability (MAP). TOSS employs two looping structures over the large-sized dataset of crystal structures to obtain an emergent library of distance distributions as the foundation for chemically intuitive understanding and then determine the OSs by minimizing a loss function for each structure based on MAP and distance distributions in the whole dataset. We apply TOSS to a dataset of over one million crystal structures, achieving a superior success rate, and use the resulting OS dataset to train a graph convolutional network (GCN) model as an alternative. Both TOSS and the GCN model are benchmarked against a curated ICSD dataset of structures with human-assigned OSs, yielding high accuracies of 96.09% and 97.24%, respectively. We expect TOSS and the ML-model-based alternative to find a wide spectrum of applications, and this work also demonstrates an encouraging example for the data-driven paradigms to explicitly compute the chemical intuition for tackling complex problems in chemistry.

## INTRODUCTION

The oxidation state (OS) is a fundamental concept that is unique to chemistry. Not only does it present a systematic and descriptive approach to understand a wide range of phenomena in chemistry,[1-3] but it also serves as a guiding landmark for chemists to pursue novel chemical compounds.[4-11] Moreover, with the rapidly increasing pervasion of machine learning (ML) techniques in solving chemistry problems[12-31] nowadays, there have been rejuvenating interests in the OS for it to serve as an intrinsic set of descriptor for in silico high-throughput materials discovery and property prediction.[32-37] Hence, this yearns for an automatic and effective computational methodology to determine the OSs for large-sized datasets of various types of atomic structures, which, however, poses a subtle challenge.

From the quantum mechanical (QM) perspective, OSs of atoms in a compound are not well defined because the electron density is global and there are no fundamental physical laws for defining a local atomic region in a compound for partitioning. Nevertheless, a rich set of partition shemes for the electron density was developed, from the classic Mulliken analysis[38] to the quantum theory of atoms in molecules by Bader[39, 40], but these schemes still lack rigorous physical justification and generally assign fractional charges to atoms in a compound, which require further classification into the integer OSs following rules with empiricism in general[41]. A similar scenario is present in the experimental determination of OSs in a compound using characterization techniques such as the X-ray absorption spectroscopy[42-44], where signals are compared with standard references and the assignment of integer OS also involves empiricism in general.

The general presence of empiricism in determining the OSs in a compound arises from the lack of rigorous definition of OS at the QM level, but this should not compromise the fundamental role of OS in chemistry because the level of scientific complexity in chemistry can require a different and emergent conceptual structure[45-47] based on concepts such as the OS and chemical bond[48] that are generally not rigorous at the QM level, and the OS as a descriptor can benefit from an immensely rich knowledge of chemistry, such as those well-documented cases showing intrinsically different catalysis by the same transition metal with different OSs[49], while the fractional charges based on various partition schemes may not be able to well characterize this knowledge[50].

On the other hand, empiricism is data-driven in nature (so is the chemical intuition), and its general presence implies a pratical approach to determine the OS, i.e., based on data, and the bond valence model (BVM)[51, 52] method well illustrates this approach. The key to applying BVM to determine the OS is the set of bond valence parameters, which was derived from the crystal structure dataset. The BVM method enables efficient determination of OSs for large-sized datasets of atomic structures, and either the QM calculations or the experimental characterizations can be formidably time-consuming for this task. However, the applicability of BVM method is greatly limited by the availability of bond valence parameters as well as the transferability of bond valence parameters to novel compounds with unusual OSs. Recently, the BERTOS model[53] and a module in Pymatgen[54] have been developed for effective and rapid predictions of OSs based on only the compositions. In addition, Mueller[55] introduced a sophisticated composition-based ML model for predicting OSs. While composition-based models enable rapid prediction, a structure-based method may resemble more the chemists' intuitive approach to assigning OSs. By analyzing local coordination environments in a structure, such method captures subtle bonding and mixed-valence scenarios that composition-based models often miss. The two approaches can be thus complementary: composition-based models offer fast and broad applicability, while a structure-based method can provide chemically intuitive OSs with interpretability.

In this work, we present a universally applicable data-driven method and the corresponding program named Tsinghua Oxidation States in Solids (TOSS) to explicitly and efficiently compute the chemically intuitive OSs in the inorganic crystal structures based on the large-sized dataset with structural information and Bayesian approach. TOSS is a fully automated computational algorithm that imitates the process of building the chemical intuition for assigning the OS. It incorporates two looping processes: (i) abstracting the distance thresholds for the analysis of local coordination environment by "learning" over all the atomic structures in the dataset repeatedly to reach converged results; (ii) determining the OSs by "practicing" over all the atomic structures in the dataset repeatedly to minimize a loss function for each structure based on only the Bayesian maximum a posteriori probability (MAP) and the distance distributions in the whole dataset. The MAP estimation is a statistical technique that combines prior information (here is the overall distribution of bond lengths and coordination environments) with observed data to find the most probable set of OSs. In our approach, minimizing the MAP-based loss function is conceptually similar to finding the lowest-energy state in a physical system (more detaled explaintions are available in the SI). This makes TOSS generally applicable to any large-sized dataset containing either existing structures or brand-new structures created by techniques such as generative models, so it is well-suited for in silico high-throughput materials discovery and property prediction. Consequently, TOSS also renders a library consisting of conceptual pictures and parameters generalized from the given dataset, including the distance distributions for all available element pairs that imply the bonding scenarios and the thus derived coordination radius for each element with a corresponding deviation that characterizes the flexibility of coordination, which can be used as chemically informative descriptors for materials discovery and property prediction. This library forms the foundation of chemically intuitive understanding toward determining the OSs in solids.

We apply TOSS to the large-sized dataset of crystal structures combining those from the legacy version (accessed on May 13$^{th}$, 2021) of Materials Project (MP)[56] and the version 1.5 of the Open Quantum Materials Database (OQMD)[57] Since these structrues lack OS labels, we perfomed a double validation by cross-referencing TOSS results with those by the BVM method. Structures with consistent OS assignments from both methods were retained, yielding a screened subset of 250,512 high-confidence entries. Using this subset, we benchmarked four graph-based and two feature-based ML models using intermediate results from TOSS (i.e., the local coordination environment) and found the simple graph convolution network (GCN) model to be the most accurate, predicting OSs with 98% accuracy. To validate both TOSS and the GCN approach, we benchmarked them against a curated ICSD dataset with human-assigned OS labels, where they demonstrated high acurraccies of 96.09% and 97.24%, respectively.To accelerate the GCN-based workflow,we further developed a link-prediction model to predict the local coordination environment from raw crystal structures, which, combined with the simple GCN model, serves as a complete ML-model-based data-driven alternative to TOSS. Both TOSS and ML-model-based alternative are available on https://github.com/yueyin19960520/TOSS, which we expect to find applications in a wide spectrum of problems that require generating OSs as the intrinsic descriptors for large-sized datasets of crystal structures. The resulting OSs and the associated library for chemically intuitive understanding are available on https://www.toss.science, providing a foundation for data-driven OS prediction and related ML applications. It is important to note, however, that OSs for many crystals can be reliably obtained from density functional theory (DFT) calculations. In this context, our approach is intended to offer an efficient alternative for large-scale applications where DFT calculations are prohibitively time-consuming.

The OS is a basic chemical concept that cannot be rigorously computed with physical laws but perfectly embodies the chemical intuition, so the data-driven paradigm introduced here for computing the OSs may serve as an exemplary paradigm for computing the chemical intuition, and this may be further employed to accelerate calculations of complex chemical systems and tackle complex problems in chemistry such as the construction of reaction network in heterogeneous catalysis.[58] This may also imply that the data-driven paradigm is a promising approach to compute the concepts emerged in the disciplines dealing with complex systems such as chemistry.

## RESULTS AND DISCUSSION

### Workflow of TOSS.

Figure 1 presents the complete workflow of TOSS, which encompasses two core looping structures to deliver the foundational library for chemically intuitive understanding and the subsequent determination of OSs.

In the first looping structure, the primary purpose is to abstract the distance thresholds from the dataset that are key parameters for defining the local coordination environment, and the threshold is defined as the longest bond length that can be counted as coordination between each pair of elements. All the thresholds are initialized as 1.5 times of the sum of Pyykkö's single-bond covalent radii[59] for each pair of elements (more discussion in Supplementary Note 1) but are converged to the emergent values from the given dataset and then should be independent of the initial guesses. The whole dataset of crystal structures is preprocessed by the "Get Structures" and "Pre-Set Features" modules (details in Supplementary Note 2) and the resulting data stream is fed to the "Digesting Structures" module, which outputs the assembly of the local coordination environment of each atomic site in the dataset.

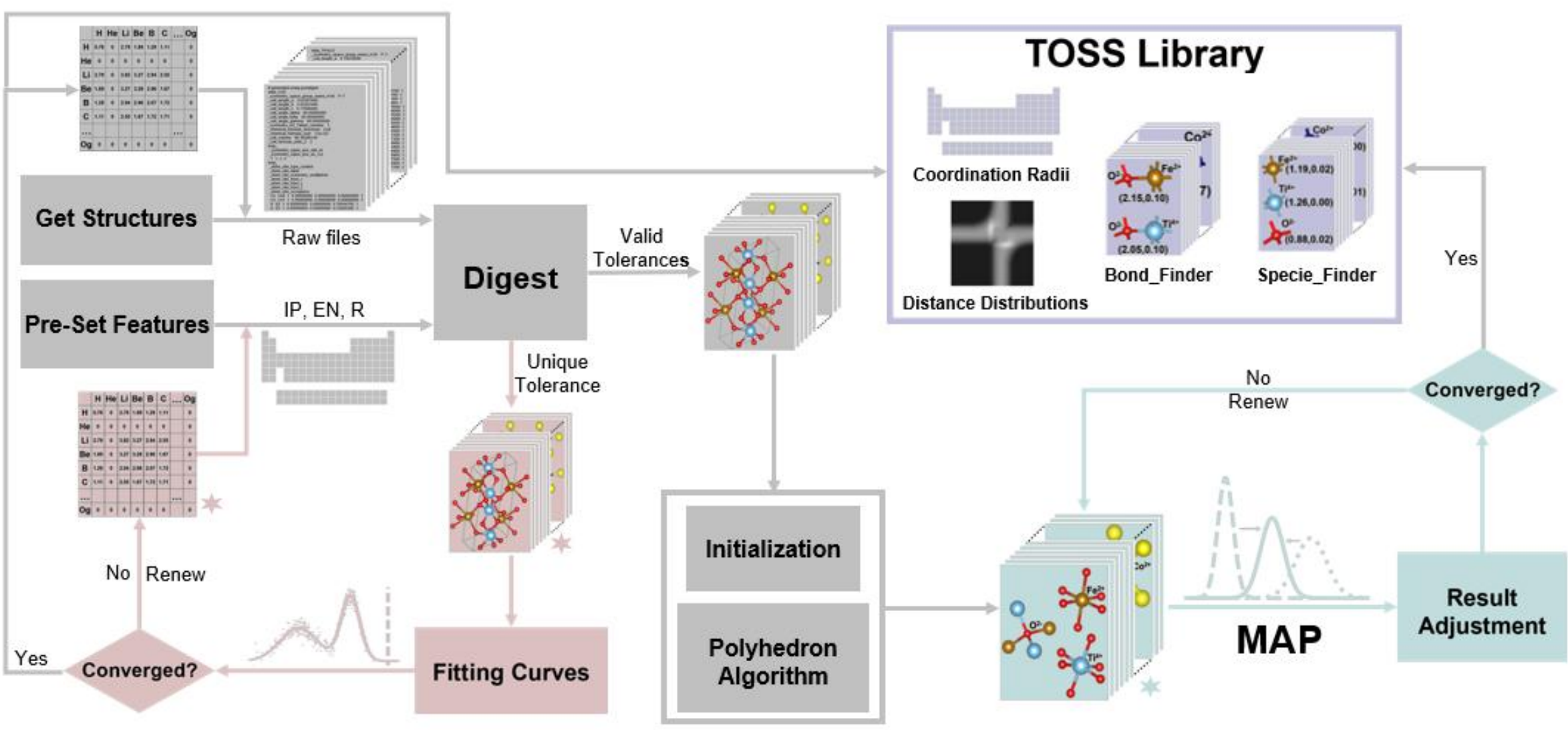

**Figure 1.** Workflow of TOSS. The subscripted stars mark the intermediate results within the looping processes.

In the “Digesting Structures” module (more details in Supplementary Note 3), for each atomic site, a sphere is first defined by the distance to its nearest neighbor multiplied by a tolerance parameter ($t$) as the radius, and within the sphere its coordination environment is then determined based on the thresholds, identifying a constituent; for each crystal structure, this is repeated for a set of $t$ values from 1.1 to 1.25 by the step of 0.01. In the first loop, only one $t$ value is chosen based on Pauling’s rule of parsimony[60] to yeild the fewest distinct kinds of constituents. However, outside this loop, all valid $t$ values, along with their distinct coordination environment sets, are provided as inputs for the second loop (more details in Supplementary Note 4). After assembling all coordination environments from the dataset, all bond lengths for each element pair are collected to generate the corresponding bond length distribution, or more formally referred to as the distance ditribution. Subsequently, each resulting distance distribution is fitted by a linear combination of two normal distribution functions in the “Fitting Curves” module, and the fitted two means ($\mu$) with corresponding deviations ($\sigma$) render a larger $\mu + 4\sigma$ as a new threshold (more details in Supplementary Note 5). By feeding the new set of thresholds back to the “Digesting Structures” module, the looping structure runs the iteration until 99.5% of the thresholds are converged, eventually forming the emergent threshold matrix for the given dataset (noting that this convergence may inherit the dataset-specific characteristics). The convergence criterion of 99.5% can be arbitrarily increased provided that the given dataset is sufficiently large (more details in Supplementary Note 6). It is worth emphasizing that, besides the thresholds, the distance distributions are also important results from the first looping process, because they are the basis for the sets of $\mu$ and $\sigma$ used as the key compoents in the formulation of the loss function in the second looping structure as discussed later.

After the first looping process, all structures with pre-set features are assembled by the "Digesting Structuers" module, carrying their valid tolerance values and corresponding sets of coordination environments. These are then fed to the "Initialization" module and "Polyhedron Algorithm" module. The "Initialization" module introduces a bond order matrix based on Pyykkö's covalent radii as a reference (more details in Supplementary Note 7), and then the "Polyhedron Algorithm" module assign the initial OSs based on the bond order matrix, Tantardini-Oganov electronegativities[61], and ionization potentials (more details in Supplementary Note 8, including four examples that illustrate the process in detail). These empirical parameters are introduced here only for generating the initial guesses that can lead to a robust convergence to the final OSs emergent from the given dataset, which should be independent of these parameters.

In the second looping structure, the Initial Guess process occurs first. At this stage, TOSS retrieves the number of distinct coordination environments ($N_t$) by varying $t$ for each structure from the the "Digesting Structures" module and includes all $N_t$ distinct coordination environments and their cooresponding OS results for all structures. For better processing these ensembles of structural information, TOSS labels each coordination bond length by the element types (ETs), coordination numbers (CNs), and OSs of the two terminal atoms (e.g., the coordination bonds formed by 6-coordinated $Fe^{3+}$ and 4-coordinated $O^{2-}$ are labeled differently from those by 4-coordinated $Fe^{2+}$ and 4-coordinated $O^{2-}$). By analysing the distance distribution of all bond lengths sharing the same label, the resulting mean and standard deviation are regarded as the emergent bond length ($\mu_{\mathrm{em}}$) and its corresponding spread ($\sigma_{\mathrm{em}}$) for each specific coordination bond type in the given dataset (more details in Supplementary Note 9). It is worth emphasizing that, at this stage, the preceding algorithms prepare an ensemble of differing sets of chemically plausible OS values for every structure determined by different tolerances, which is the input for the second looping structure to determine the most probably OS values, *i.e.*, a single set of optimized OS assignments for each structure. Note that, in preparing the ensemble of differing sets of chemically plausible OS values, we develop a set of algorithms in the "Polyhedron Algorithm" module including a Resonance method to make up certain missing but chemically plausible OS values, particularly for the cases containing alkali or alkaline earth elements, and more technical details with illustrative examples can be found in Supplementary Note 8.

In order to evaluate every chemically plausible OS set, we derive a loss function based on MAP in Bayesian statistics, which provides estimation of unobserved quantity on the basis of the whole dataset. The loss function for each structure in the dataset bears the form derived from MAP as follows (the derivation is provided in Supporting Information)

$$\mathrm{LOSS} = \sum_{L_i}\left[\frac{(L_i-\mu_i)^2}{2\sigma_i^2} - \log\left(\frac{\Delta_L}{\sqrt{2\pi}\sigma_i}\right)\right] - \sum_i n_i \log\left[p_i\left(=\frac{N(\mathrm{ETs},\mathrm{CNs},\mathrm{OSs})}{N(\mathrm{ETs},\mathrm{CNs})}\right)\right]$$

where $i$ is the label of bond type identified by ETs, CNs, and OSs of terminal atoms; $L_i$ is the bond length of a coordination bond with the label $i$; $\mu_i$ and $\sigma_i$ are the emergent bond length and its deviation from the dataset; $\Delta_L$ is the integral interval (taken as 0.001Å, which can be regarded as the threshold for distinguishing different bond lengths; this value reflects the typical resolution of bond distances obtained from modern X-ray diffraction (XRD) measurements and DFT calculations); $n_i$ is the number of bonds sharing the label $i$ in a structure; $p_i$ is the occurring probability of a specific set of OSs for a certain set of ETs and CNs in the dataset and can be calculated as the ratio in the following parenthesis. The first term sums over all $L_i$ available in a structure and characterizes how likely a bond belongs to a label, while the second term sums over all label $i$ identified in a structure and evaluates how likely a bond type corresponds to a set of OSs based on the dataset. The loss function here is the negative logarithm of MAP, so it is to be minimized for the best estimation of OSs for each structure based on the distributions in the whole dataset. The MAP estimation combines the prior information—derived from the global distributions of bond lengths and OS occurrences—with the likelihood of the observed data for each structure. Minimizing the negative log of this combined probability effectively selects the OS assignment that best fits both the local structural data and the overall trends in the dataset, similar to the optimization for a system's most stable, lowest-energy configuration (more detaled explaintions are available in the SI).

Subsequently, in the second looping process, the "Result Adjustment" module varies the OSs across all tolerances subject to only the integer OS and neutrality constraints, and this can result in new sets of $\mu_i$, $\sigma_i$, and $p$, with which TOSS re-evaluates the MAP-based loss function for each structure to identify the single set of OSs and corresponding coordination enviornment leading to the lowest loss value (more details in Supplementary Note 10). Thus, the second looping structure runs the iteration with the "Result Adjustment" module until 99.5% of the structures' results for the entire dataset do not change, delivering the final $\mu_{\mathrm{em}}$, $\sigma_{\mathrm{em}}$, and emergent OSs. This well resembles the self-consistent field (SCF) approach, because the MAP-based loss function for each structure (like the one-particle equation) depends on the distribution of OSs in the whole dataset (like the mean field) via the set of $\mu_i$, $\sigma_i$, and $p$. It is worth emphasizing that the MAP-based loss function is iteratively evaluated for all the structures, because whenever there is a change in the OS value(s) in the dataset, the MAP-based loss functions of all the structures are updated. Besides, the optimization of MAP-based loss functions does not changes the OSs but selects the most probable set of OSs for every structure from the ensemble of differing sets of chemically plausible OSs provided by various algorithms in TOSS.

**Library for Chemically Intuitive Understanding.**

In the first looping structure, the resulting distance distribution for each element pair from the large dataset provides a manifest foundation for understanding the chemical coordination. As an obvious illustration, Figure 2a shows the plot of distance distribution for the O-Al pair that demonstrates two peaks around 1.757 Å and 1.895 Å, which mainly corresponds to two types of coordination bonds in the dataset (containing 235,632 O-Al bonds) formed by O with 4-coordinated Al and 6-coordinated Al, respectively, and this well conforms to the chemically intuitive understanding of O-Al bond lengths. A more complicated example is the distance distribution for the O-V pair shown in Figure 2b, and its multiple peaks are a result of mixing a few different types of coordination bonds in the dataset (containing 424,430 O-V bonds) including the ones formed by O with 4-coordinated $V^{5+}$ (1.75-1.77 Å depending on the CN of bonded O), 5-coordinated $V^{5+}$ (1.84-1.92 Å), and 6-coordinated $V^{5+}$ (1.89-1.95 Å), $V^{4+}$ (1.97-2.02 Å) and $V^{3+}$ (1.99-2.07 Å), which exactly lays the foundation of well-educated chemical intuition for the different OSs of V atoms in the crystal structures.

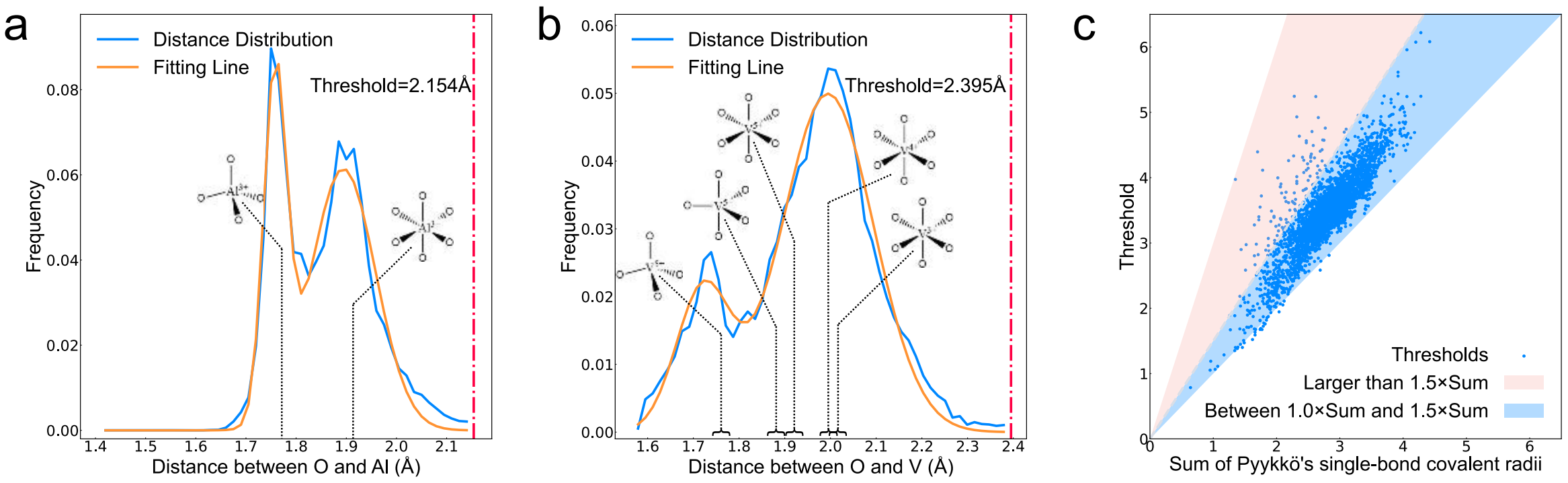

**Figure 2.** Distance distributions (the frequency is normalized by the total number of pairs in the dataset) between **(a)** O and Al, **(b)** O and V, fitted by the linear combination of two normal distribution functions, which renders the larger $\mu + 4\sigma$ as the distance threshold for coordination between each element pair. The insets mark the main types of constituents in the distributions. **(c)** Correlation between the converged distance thresholds and the sums of Pyykkö's single-bond covalent radii.

To determine a distance threshold for coordination between each element pair from the distance distribution, we adopt that the bond length distribution for any coordination bond type follows a normal distribution owing to a large number of variables tweaking the bond length in the large dataset of crystal structures, so we fit each distance distribution with a linear combination of normal distribution functions. The number of normal distribution functions used for fitting should depend on the number of coordination bond types (constituents) for each element pair, but in practice, we found that a linear combination of two independent normal distribution functions is sufficiently robust (as exemplified by Figures 2a and 2b) for fitting all the distance distributions to obtain just the thresholds, i.e., the maximum bond length for forming a coordination bond (regardless of its type) between an element pair. The fitting function form is thus expressed as follows,

$$f(x) = A_1 \exp\left[-\frac{(x-\mu_1)^2}{2\sigma_1^2}\right] + A_1 \exp\left[-\frac{(x-\mu_2)^2}{2\sigma_2^2}\right]$$

where $A_i$, $\mu_i$, and $\sigma_i$ are the coefficient, mean, and deviation for each normal distribution function, respectively. The distance threshold for coordination between each pair of elements $i$ and $j$ is then obtained as follows,

$$T_{ij} = \max(\mu_1 + 4\sigma_1, \mu_2 + 4\sigma_2) \equiv \mu + 4\sigma$$

where $4\sigma_i$ is taken to include 99.99% of the bond length distribution and the terms in the obtained threshold are denoted as $\mu$ and $\sigma$ for simplicity. Both the fitting curves and the obtained thresholds are illustrated in Figures 2a and 2b. It is noteworthy that the O-Al coordination bond length has a smaller fitted deviation (or spread) $\sigma$ of 0.065 Å than the $\sigma$ of 0.100 Å for the O-V coordination bond length, implying that the O-Al coordination bond is more rigid than the O-V coordination bond, which well conforms to the chemical intuition. More details of fitting are provided in Supplementary Note 5, and all the distance distributions with fitting results are available on https://www.toss.science in the form of a clickable periodic table for easy access.

Figure 2c plots all the converged distance thresholds for coordination between element pairs against the corresponding sums of Pyykkö's single-bond covalent radii, which are adopted to provide the initial guesses of thresholds in TOSS. The positive correlation shown in Figure 2c justifies the use of Pyykkö's radii for intial guesses, but the widely spread distribution of thresholds implies that they are emergent from the given dataset and should be independent of initial guesses. We tested the use of 1.5 times the sum of Pyykkö's radii as a simple threshold set and found that this results in 15.69% different coordination environments in the dataset. This underscores the importance of using self-consistent thresholds for defining the coordination environment.

With the obtained distance thresholds for coordination between element pairs, we can employ the properties of normal distribution functions to naturally derive the coordination radius and the associated spread for each element. Because the coordination bond length distribution $\boldsymbol{L_{ij}}$ for any pair of elements $i$ and $j$ is adopted to be a normal distribution $\boldsymbol{\mathcal{N}_{ij}}$ as $\boldsymbol{L_{ij}} \sim \boldsymbol{\mathcal{N}_{ij}}(\mu_{ij}, \sigma_{ij})$, we further adopt that the atomic radius distribution $\boldsymbol{R_i}$ for each element $i$ to form the coordination also follows a normal distribution as $\boldsymbol{R_i} \sim \boldsymbol{\mathcal{N}}(\mu_i, \sigma_i)$, then $\boldsymbol{L_{ij}}$ is simply the convolution of $\boldsymbol{R_i}$ and $\boldsymbol{R_j}$ as follows (the derivation is available in Supplementary Note 11),

$$\boldsymbol{L_{ij}} \sim \boldsymbol{\mathcal{N}_{ij}}(\mu_{ij}, \sigma_{ij}) = \boldsymbol{R_i} \otimes \boldsymbol{R_j} = \frac{1}{\sqrt{2\pi(\sigma_i^2 + \sigma_j^2)}} \exp\left\{-\frac{[x - (\mu_i + \mu_j)]^2}{2(\sigma_i^2 + \sigma_j^2)}\right\}$$

so the coordination bond length and its spread are expressed as

$$\mu_{ij} = \mu_i + \mu_j$$

$$\sigma_{ij} = \sqrt{\sigma_i^2 + \sigma_j^2}$$

where $\mu_i$ is the coordination radius for element $i$ and $\sigma_i$ is the corresponding spread that characterizes the flexibility for element $i$ to form a coordination bond. Here, it is more chemically intuitive to specify the index $i$ to identify the specific form of the bonding atom with the CN and OS in addition to the ET, because $\mu_i$ and $\sigma_i$ should depend on all of them. Thus, by solving the two sets of multivariable equations based on $\mu_{ij}$ and $\sigma_{ij}$, respectively, we obtain the values of $\mu_i$ and $\sigma_i$ for all available forms of each element with different values of CN and OS.

Figure 3 lists the coordination radius and spread for the most frequent form (labeled by the CN and OS, excluding the trivial alloy forms with the OSs of zero) of each element. The results are generally consistent with the chemical intuition, and it is woth noting that the spreads of the cationic forms are commonly less than those of the anionic forms, implying that the cations are less flexible to form the coordination bonds than the anions. The values of $\mu_i$ and $\sigma_i$ for all forms of each element are available on https://www.toss.science. These are valuable data for building the chemical intuition, such as making quick educated guesses of OSs and local coordinations in a crystal structure, and more importantly, they can be used as chemically informative descriptors for materials discovery and property prediction.

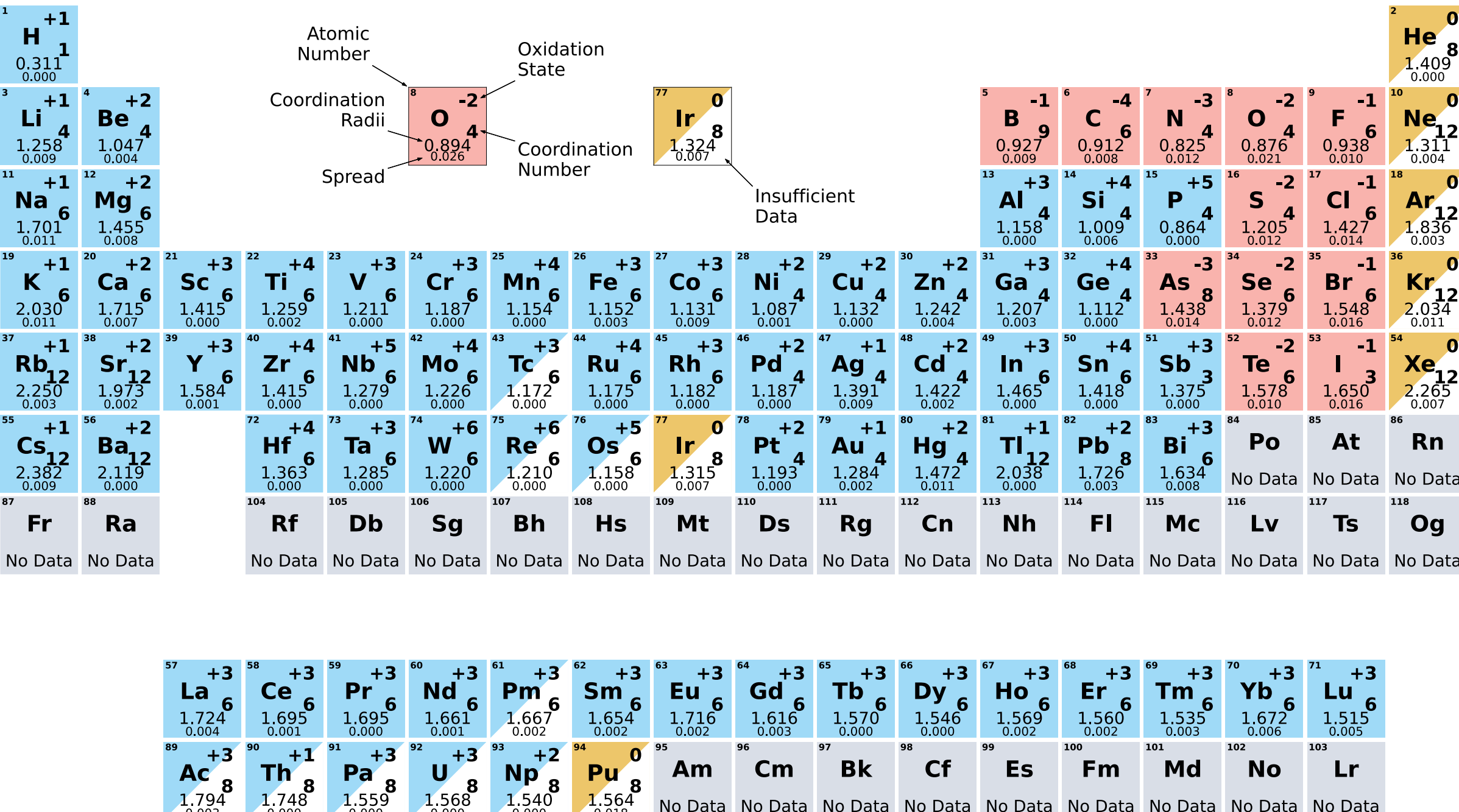

**Figure 3.** Periodic Table of the most frequent element form (labeled by OS and CN) in the dataset (excluding the trivial alloy forms for metals) with the corresponding coordination radius and spread. The elements with less than 20,000 occurences in all labeled coordination bonds are marked with insufficient data.

**OS Results by TOSS.**

For the combined large dataset of 1,147,168 crystal structures obtained from the MP and OQMD databases, TOSS successfully assigns OS values for 1,114, 330 crystal structures, i.e., a success rate of 97.14%, and this is much superior to that of 33.57% by BVS (an alternative name of BVM, as implemented in the pymatgen package with default parameters; more details are available in Supplementary Note 12), as shown in Figure 4a. The success rate by TOSS does not reach 100% because of two occasions. First, the initial assignment of OSs in the "Initialization" and "Polyhedron Algorithm" modules fails to work for a small portion of structures, because they have too complicated coordination scenarios to successfully assign the intial guesses of OSs. Second, for the sake of computational cost, we adopt a convergence criterion of 99.5% and the rest 0.5% of results are marked as unsuccessful, among which certain coordination bond types have too few cases in the dataset to deliver effective convergences. In fact, a larger dataset can lead to faster convergences in general and thus a higher computational efficiency. Therefore, TOSS may achieve a higher success rate for assigning OS values given a larger dataset. Furthermore, it is worth emphasizing that TOSS works universally because it relies on only the large-sized dataset and Bayesian statistics, in stark contrast to BVS that relies on the availability of empirical parameters.

BVS successfully assigns OS values for only 385,067 crystal structures in the dataset, 373,177 of which TOSS also does so. Among these results, TOSS and BVS agree with each other on the OS values of 250,512 structures but give different results for the rest 122,665 structures, as shown in Figure 4a. Importantly, 33,746 of these structures are alloys (an alloy is defined here as the structure composed of only metal elements and in which the differences of electronegativity between bonded metal atoms are less than 1.0), which should be excluded from this comparison, because TOSS assigns the OS of zero to all metal components in alloys, whereas BVS cannot assign zero OS values due to the lack of corresponding parameters. Nevertheless, for the 88,919 structures (excluding the alloys) with different OS values assigned by TOSS and BVS, there are no definite simple rules to evaluate them, and we list 100 example structures randomly selected among them on https://www.toss.science/examples along with their OS values assigned by TOSS and BVS as well as magnetic moments and Bader charges obtained from density functional theory calculations for manual evaluation. Also, to illustrate the comparative performance of TOSS, BVS, and BERTOS, we include in the Supplementary Notes 13 a table highlighting 10 key examples where the three methods differ significantly.

Additionally, we apply TOSS to the CIF-parsed and OS-labeled entries in the ICSD dataset (excluding structures with partial occupancies or missing atoms that prevent further processing). TOSS successfully assigns OS values to 79,146 structures, achieving an accuracy rate of 96.09% compared to human-assigned OS values, as detailed in the confusion matrix in Figure 4b. While TOSS performs exceptionally well overall, its few discrepancies primarily arise from element-specific issues. For instance, TOSS assigns $-5$ to B in borides and $-4$ to C in certain carbides, whereas the ICSD dataset often labels these atoms with an OS of 0, which may be arguable. Similarly, the OS of +8 assigned to 51 atoms of Os, Ru, or Xe shows a 21.6% disagreement with ICSD labels, likely due to their atypical local coordination environments. Despite these minor limitations in extreme OSs, TOSS demonstrates high reliability for the vast majority of cases, reinforcing its suitability for large-scale automated OS assignment where manual validation is practically infeasible.

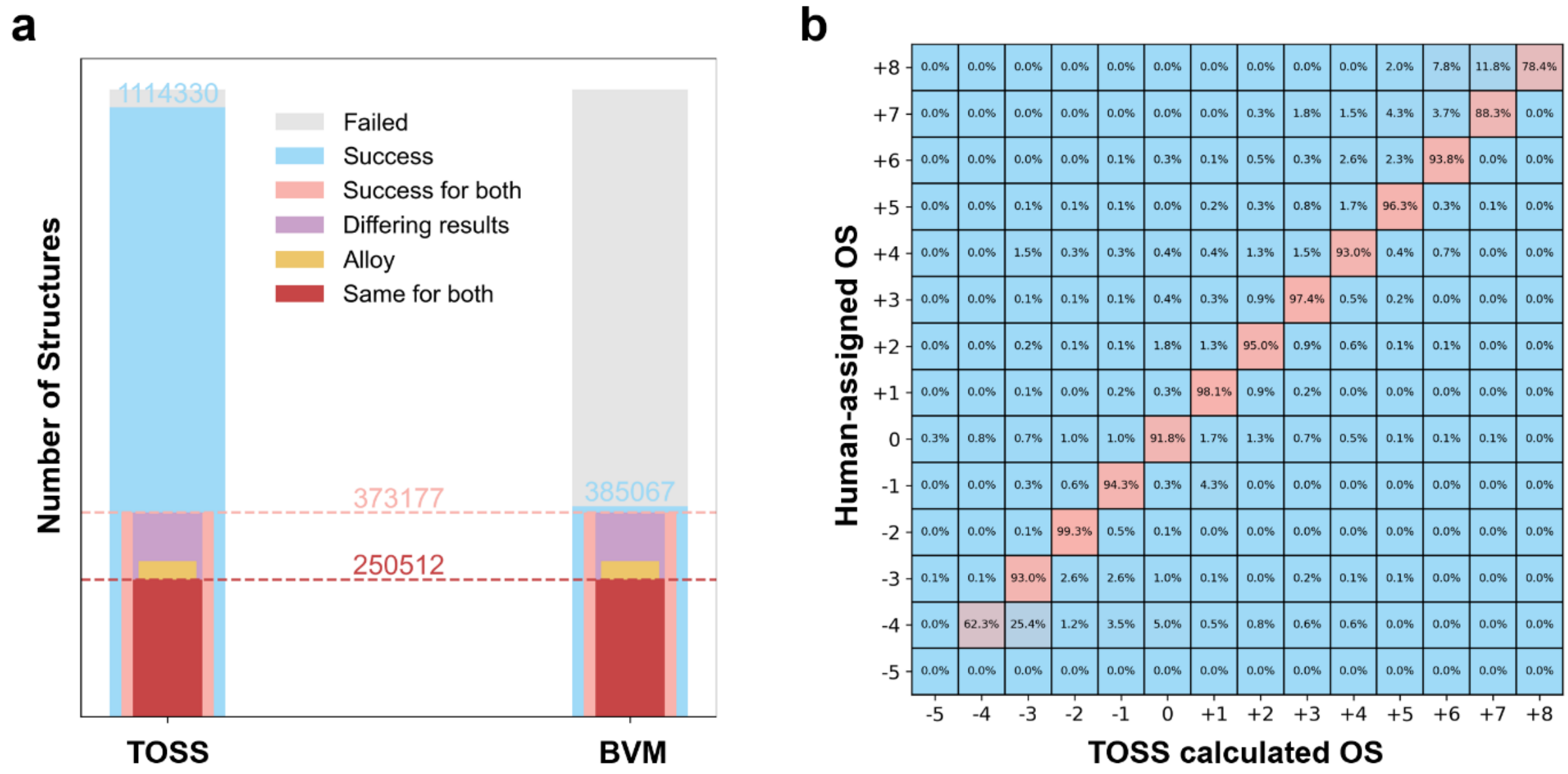

**Figure 4.** (a) Comparison of success rates and OS results by TOSS with those by BVS. (b) Confusion matrix evaluating TOSS calculated OSs and human-assinged OSs in the ICSD dataset.

## Current limitations of TOSS.

Before proceeding to the applications of ML models, it is worth noting the current limitations of TOSS. In the design principle underlying TOSS, the coordination environment can be defined by not only the CN and coordinating elements but also the coordination bond lengths. This definition of coordination environment is capable of distinguishing subtle cases exemplified in Figure 2a,b, which demonstrates that varying CNs and/or OSs can be clearly identified by different coordination bond lengths. However, TOSS fails to distinguish subtle cases such as Sb(III) and Sb(V) in $Cs_2SbCl_6$ with differing Sb(III)-Cl and Sb(V)-Cl bond lengths. This highlights a practical limitation of TOSS in its current status. In order to boost the computational efficiency of TOSS over large datasets for affordable tests in the current development of TOSS as a proof of concept, we adopted the simplifying assumption that we ignored the subtle variations of coordination bond lengths in the coordination environment, and this is the key culprit for the failure in distinguishing Sb(III) and Sb(V) in $Cs_2SbCl_6$. To address this subtlety, we have implemented a patch for TOSS to include the consideration of different coordination bond lengths when their difference is larger than a threshold of 0.05 Å, which enables TOSS to successfully identify Sb(III) and Sb(V) in $Cs_2SbCl_6$ while only slightly increases the computational cost.

Besides, another limitation of TOSS can be demonstrated by the 100 examples on https://www.toss.science/examples. These examples show that TOSS generally assigns chemically intuitive OSs but fails in some cases, which we attribute to insufficient data in the dataset (e.g., the OS of lanthanide/actinide by TOSS is less chemically intuitive because the dataset contains much less structures containing lanthanide/actinide than other elements). Hence, we expect that the capability of TOSS can be systematically improved by including more data for every element pair in the future development. In addition, as a proof-of-concept work, we did not conduct data cleaning of the structures from MP and OQMD, and we plan to update both the size and the quality of the dataset for TOSS in the future development.

Also, TOSS has intrinsically practical limitations that become evident when comparing with ML approaches. TOSS uses an ensemble of rule-based algorithmic methods, which are robust across most cases but can struggle with ambiguous or complex structures. In contrast, ML approaches such as the graph convolutional network (GCN) model can predict OSs directly from atomic and local coordination features, offering better feasibility, efficiency and scalability. However, it is important to emphasize that TOSS can serve as the essential cornerstone for ML approaches, providing high-quality datasets of OSs required for training, which dictate the reliability of ML approaches.

In addition to these factors, it is important to acknowledge that the iterative convergence of the distance distributions—which underpins the emergent threshold matrix—can also be influenced by dataset-specific constraints. Since the fitting process of these distributions relies on the bond length data available in the present dataset, any overrepresentation or underrepresentation of certain bonding environments may introduce bias into the converged thresholds. Thus, the results may reflect the idiosyncrasies of the present dataset rather than a universally applicable chemical standard, while this bias can be minimized by a large-sized dataset as we used in this work with over one million crystal structures.

**ML Models.**

The TOSS-calculated OS results of 1,114,330 crystal structures compose a large OS dataset for training ML models, but for the assured quality of data and eliminate the impact of human-assigned OS results in the ICSD dataset (used as the external test set), we take only the OS reults of 250,512 structures that BVS and TOSS give the same values to form the dataset, which is, according to standard ML practice, split internally into the training set, the validation set, and the test set by the rario of 3:1:1.

The graph convolutional network (GCN) has been successfully applied to various prediction tasks in chemistry based on the atomic structures[15, 62-67], so we benchmark four GCN models for predicting the OSs of crystal structures, including the simple GCN[64, 68, 69], the graph attention network (GAT)[70], Attentive FP[71], and the message passing neural network (MPNN) models[72]. In addition, the ensemble learning is considered as one of the state-of-the-art methods to solve a prediction problem, so we also include the feature-based random forest (RF)[73] and XGBoost[74] models for benchmark. The input of GCN models includes the features of atoms and bonds as well as the link relationships between atoms. The input features of atoms are composed of the element properties used in the "Initialization" module of TOSS (the atomic number, Pyykkö's covalent radii, Tantardini-Oganov electronegativities, and ionizaition potentials) and also the output information about the coordination environment by TOSS (the coordination number and the coordinating atoms with their properties). The input feature of bond is just the bond length. The input of feature-based models takes only the features of atoms (more details are available in the section of Details about ML Models in the SI).

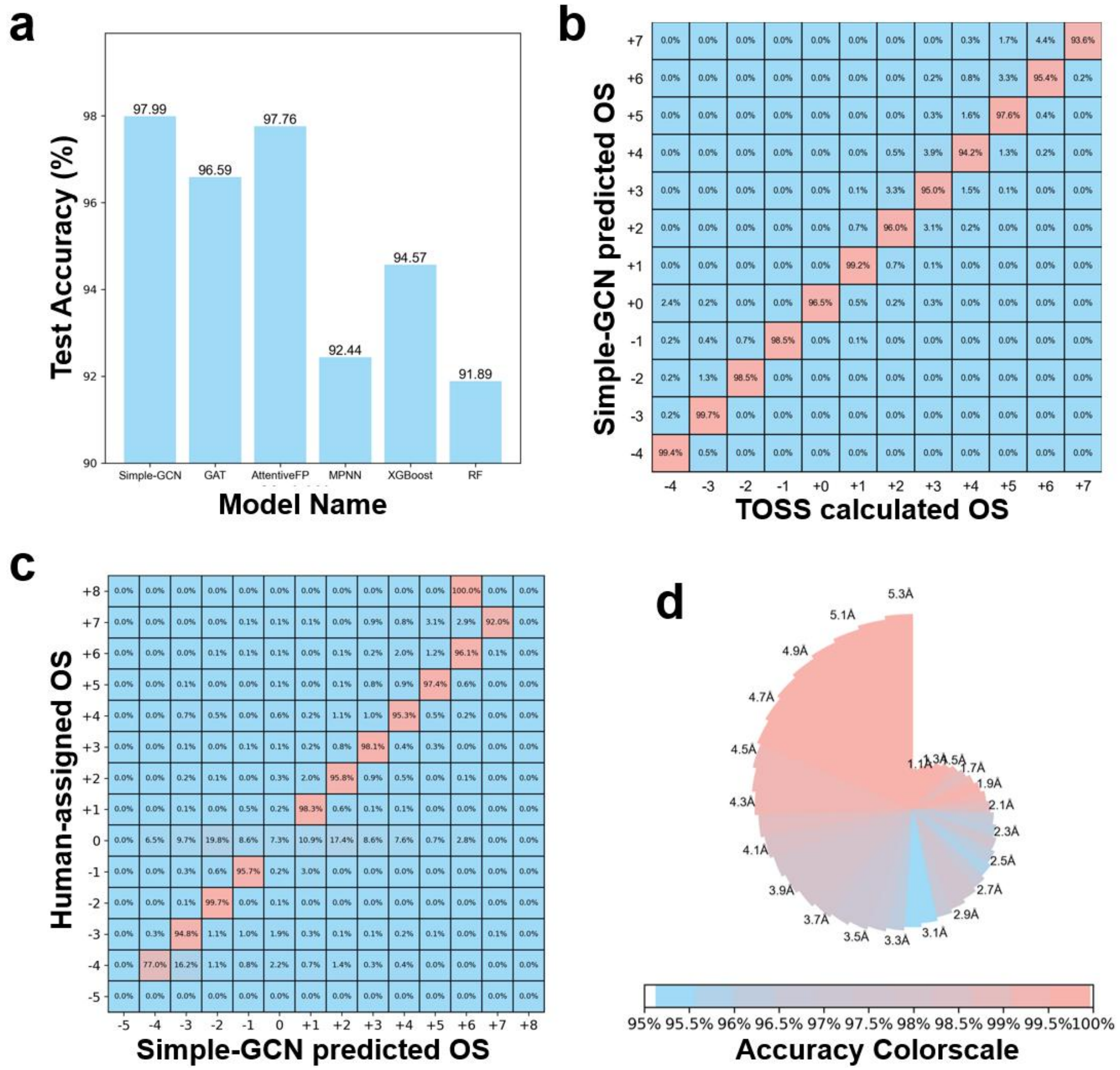

**Figure 5.** (a) Performances of four graph-based models and two feature-based models for predicting OSs. (b) Confusion matrix evaluating the TOSS calculated OSs and simple GCN predicted OSs. (c) Confusion matrix evaluating the simple GCN predicted OSs and human-assinged OSs in the ICSD dataset. (d) Accuracy of link prediction model for predicting coordination grouped by bond lengths.

Figure 5a shows that the simple GCN model delivers the best accuracy of 97.99% for predicting the OSs in solids, and the accuracies by GCN models are generally better than those by feature-based models, among which the XGBoost model is better. The performance of the simple GCN model can be further assessed by the confusion matrix shown in Figure 5b, which highlights the model's generally high accuracies for OS prediction categorized by the OS values, although the prediction of positive OSs is slightly less than that of negative OSs. When adjudicating 88,919 structures with TOSS-BVS disagreements, the GCN model showed comparable alignment with both methods (77.65% TOSS agreement vs 77.00% BVS agreement). However, TOSS maintains a clear advantage in applicability (97% success rate vs 34% by BVS) due to its parameter-free, data-driven design that automatically extracts chemically relevant bond-length distributions. The GCN model demonstrates exceptional performance on the ICSD dataset, achieving 97.24% accuracy across 85,526 curated structures. While matching the accuracy of composition-based methods like BERTOS, it extends seamlessly to a much larger and more structurally complex dataset. This result underscores the model's strong generalization to real-world materials and its superior scalability to diverse crystal structures without requiring manual parameter tuning. As shown in Figure 5c, the model's design intentionally excludes predictions for +8 or −5 OSs due to insufficient training data. In practice, all +8 elements (Os, Ru, and Xe) are predicted as +6 and no −5 OS appears in the labeled ICSD structures. The relatively lower accuracy for zero OS mainly reflects the same element-specific issues discussed earlier in the TOSS confusion matrix result. Among the 2,419 cases with the OS of 0 labeled by ICSD, 2,281 (94.30%) come from H, C, N, and Si, such as in $B_2H_6$, $Al_4C_3$, $Fe_3N$, $Al_2EuSi_2$, which are also arguable (all these mismatch cases with ICSD-labeled zero OS are listed on our GitHub page). This also shows that our GCN model is well-aligned with TOSS that provides its training data.

The trained simple GCN model above requires the input of information about the coordination environment output by TOSS, and to leverage the capability of ML models, we further developed a link prediction model to accurately predict the required information about the coordination environment directly from the raw data of crystal structures, which delivers an accuracy of 97.77%. This is achieved by introducing a heterogeneous-graph-based GCN model, which is inspired by the approach in TOSS that the ditributions of both atomic coordination radii and bond lengths can be approximated as Gaussian distributions. This allows for the abstraction of bonds as nodes within the graph, thereby facilitating the information aggregation algorithm to acquire the bond entity within the GCN architecture[75, 76] (more details are provided in Supporting Information). Figure 4d demonstrates the model's exceptional accuracy for predicting the coordination environment directly from raw crystal structure data with a wide spectrum of coordination bond lengths. Consequently, the integration of the link prediction model with the trained simple GCN model enables the direct prediction of OSs from the input crystal structures. This provides an alternative to TOSS entirely based on ML models, which is also a data-driven approach.

## CONCLUSION

To conclude, we establish a data-driven paradigm with its implementation as TOSS for determining the chemically intuitive OSs in inorganic crystal stuctures *via* two looping structures based on the large-sized dataset and Bayesian MAP. The first looping structure abstracts a set of distance thresholds for the analysis of coordination between element pairs, which is converged over the dataset and thus can be considered as emergent knowledge from the dataset and independent of empirical parameters introduced for initial guesses. By varying all the OSs in every possible way, the second looping structure then repeatedly minimizes a loss function based on the Bayesian MAP for each structure, which is constructed with the information about all the local coordination environments in the whole dataset that are obtained using the emergent thresholds from the first looping structure, and the set of minimized loss functions converged over the dataset renders the final OS results that can also be considered as emergent properties of the dataset. Therefore, TOSS is a universally applicable method for automatic and effective determination of OSs in the large-sized datasets of crystal structures. The finalized TOSS framework and the pre-trained GCN models can be directly applied to individual crystal structures, thereby extending their utility from high-throughput screening to targeted, standalone analyses.

Additionally, TOSS delivers a foundational library for chemically intuitive understanding. This includes the distance distributions between element pairs, which provide manifest foundations for understanding the coordination scenarios, and the thus derived coordination radius for each element with a corresponding spread based on the convolution of Gaussian distributions, which characterizes the element's capability and flexibility for coordination in crystal structures, respectively. Moreover, TOSS delivers a superior success rate of 97.14% for assigning OSs for the dataset combining MP and OQMD with more than 1 million crystal structures, and the OS results compose a suitable basis for benchmarking and training ML models. Thus, we identify the GCN models to be accurate for predicting the OSs and develop a heterogeneous-graph-based GCN model to predict the coordination environment from crystal structures and a simple GCN model to predict the OSs from the coordination environment, so the two ML models combine to serve as an alternative data-driven paradigm. Both TOSS and its GCN variant are benchmarked against a curated ICSD dataset with human-assigned OSs, yielding high accuracies of 96.09% and 97.24%, respectively, and many of the ICSD-labeled OSs in the mismatch cases may be arguable. We expect our TOSS and ML-model-based alternative to find applications in a wide spectrum of problems, serving as an automatic and effective tool to generate OSs as the intrinsic descriptors for large-sized datasets of crystal structures.

Moreover, the data-driven paradigms developed here, i.e., TOSS and the ML-based approach, present a type of effective methodologies to explicitly compute the OS that embodies the chemical intuition but cannot be computed with well-defined physical laws, and the effectiveness may arise from that the chemical intuition is based on experience and is thus data-driven in nature. Therefore, this work demonstrates an encouraging example for developing effective methodologies to explicitly compute the chemical intuition, and the data-driven paradigms may be further employed to develop automatic and effective methods for computing other components in the conceptual structure of chemistry, including the bond order, the Lewis structure, and the drawing of reaction mechanisms, which may serve as powerful tools to tackle a wide spectrum of complex problems in chemistry and relevant displines.

**Data availability**

The whole dataset in this work is available on https://www.toss.science. The TOSS code and ML models are available on https://github.com/yueyin19960520/TOSS. The Derivation of Loss Fuction Based on MAP, Details about ML Models, and Supplementary Notes 1-13 are available in the ESI.

**Corresponding Author**

*Corresponding author. E-mail: haixiao@tsinghua.edu.cn

**Acknowledgements**

We are grateful to Prof. Jun Li for insightful discussions. This work was supported by National Key Research and Development Project (2022YFA1503000), National Natural Science Foundation of China (92261111), and the NSFC Center for Single-Atom Catalysis (22388102). We are also grateful to the Center of High-Performance Computing at Tsinghua University for providing computational resources.

**REFERENCES**

1. P. Karen, Oxidation state, a long-standing issue!, *Angew. Chem. Int. Ed. Engl.*, 2015, **54**, 4716-4726.
2. A. Walsh, A. A. Sokol, J. Buckeridge, D. O. Scanlon and C. R. A. Catlow, Oxidation states and ionicity, *Nat. Mater.*, 2018, **17**, 958-964.
3. J. Guzman and B. C. Gates, Catalysis by supported gold: correlation between catalytic activity for CO oxidation and oxidation states of gold, *J. Am. Chem. Soc.*, 2004, **126**, 2672-2673.
4. S. P. Green, C. Jones and A. Stasch, Stable magnesium (I) compounds with Mg-Mg bonds, *Science*, 2007, **318**, 1754-1757.
5. S. Riedel and M. Kaupp, The highest oxidation states of the transition metal elements, *Coord. Chem. Rev.*, 2009, **253**, 606-624.
6. S. Krieck, H. Görls and M. Westerhausen, Mechanistic elucidation of the formation of the inverse Ca (I) sandwich complex [(thf) 3Ca (μ-C6H3-1, 3, 5-Ph3) Ca (thf) 3] and stability of aryl-substituted phenylcalcium complexes, *J. Am. Chem. Soc.*, 2010, **132**, 12492-12501.
7. D. Samanta and P. Jena, Zn in the +III oxidation state, *J. Am. Chem. Soc.*, 2012, **134**, 8400-8403.
8. G. Wang, M. Zhou, J. T. Goettel, G. J. Schrobilgen, J. Su, J. Li, T. Schloder and S. Riedel, Identification of an iridium-containing compound with a formal oxidation state of IX, *Nature*, 2014, **514**, 475-477.
9. J. Lin, S. Zhang, W. Guan, G. Yang and Y. Ma, Gold with +4 and +6 Oxidation States in AuF(4) and AuF(6), *J. Am. Chem. Soc.*, 2018, **140**, 9545-9550.
10. W. L. Li, T. T. Chen, W. J. Chen, J. Li and L. S. Wang, Monovalent lanthanide(I) in borozene complexes, *Nat. Commun.*, 2021, **12**, 6467.
11. J. T. Boronski, A. E. Crumpton, L. L. Wales and S. Aldridge, Diberyllocene, a stable compound of Be (I) with a Be–Be bond, *Science*, 2023, **380**, 1147-1149.
12. J. P. Janet and H. J. Kulik, Predicting electronic structure properties of transition metal complexes with neural networks, *Chem. Sci.*, 2017, **8**, 5137-5152.
13. J. P. Janet, L. Chan and H. J. Kulik, Accelerating Chemical Discovery with Machine Learning: Simulated Evolution of Spin Crossover Complexes with an Artificial Neural Network, *J. Phys. Chem. Lett.*, 2018, **9**, 1064-1071.
14. S. Lu, Q. Zhou, Y. Ouyang, Y. Guo, Q. Li and J. Wang, Accelerated discovery of stable lead-free hybrid organic-inorganic perovskites via machine learning, *Nat. Commun.*, 2018, **9**, 3405.
15. S. Back, J. Yoon, N. Tian, W. Zhong, K. Tran and Z. W. Ulissi, Convolutional Neural Network of Atomic Surface Structures To Predict Binding Energies for High-Throughput Screening of Catalysts, *J. Phys. Chem. Lett.*, 2019, **10**, 4401-4408.
16. M. Zhong, K. Tran, Y. Min, C. Wang, Z. Wang, C. T. Dinh, P. De Luna, Z. Yu, A. S. Rasouli, P. Brodersen, S. Sun, O. Voznyy, C. S. Tan, M. Askerka, F. Che, M. Liu, A. Seifitokaldani, Y. Pang, S. C. Lo, A. Ip, Z. Ulissi and E. H. Sargent, Accelerated discovery of CO(2) electrocatalysts using active machine learning, *Nature*, 2020, **581**, 178-183.
17. L. Chanussot, A. Das, S. Goyal, T. Lavril, M. Shuaibi, M. Riviere, K. Tran, J. Heras-Domingo, C. Ho and W. Hu, Open catalyst 2020 (OC20) dataset and community challenges, *ACS Catal.*, 2021, **11**, 6059-6072.
18. K. M. Jablonka, D. Ongari, S. M. Moosavi and B. Smit, Using collective knowledge to assign oxidation states of metal cations in metal-organic frameworks, *Nat. Chem.*, 2021, **13**, 771-777.
19. S. Lu, Q. Zhou, Y. Guo and J. Wang, On-the-fly interpretable machine learning for rapid discovery of two-dimensional ferromagnets with high Curie temperature, *Chem*, 2022, **8**, 769-783.
20. R. Tran, J. Lan, M. Shuaibi, B. M. Wood, S. Goyal, A. Das, J. Heras-Domingo, A. Kolluru, A. Rizvi and N. Shoghi, The Open Catalyst 2022 (OC22) dataset and challenges for oxide electrocatalysts, *ACS Catal.*, 2023, **13**, 3066-3084.
21. S. Pablo-Garcia, S. Morandi, R. A. Vargas-Hernandez, K. Jorner, Z. Ivkovic, N. Lopez and A. Aspuru-Guzik, Fast evaluation of the adsorption energy of organic molecules on metals via graph neural networks, *Nat. Comput. Sci.*, 2023, **3**, 433-442.

22. N. J. Szymanski, B. Rendy, Y. Fei, R. E. Kumar, T. He, D. Milsted, M. J. McDermott, M. Gallant, E. D. Cubuk, A. Merchant, H. Kim, A. Jain, C. J. Bartel, K. Persson, Y. Zeng and G. Ceder, An autonomous laboratory for the accelerated synthesis of novel materials, *Nature*, 2023, **624**, 86-91.
23. J. Lan, A. Palizhati, M. Shuaibi, B. M. Wood, B. Wander, A. Das, M. Uyttendaele, C. L. Zitnick and Z. W. Ulissi, AdsorbML: a leap in efficiency for adsorption energy calculations using generalizable machine learning potentials, *npj Comput. Mater.*, 2023, **9**.
24. A. Merchant, S. Batzner, S. S. Schoenholz, M. Aykol, G. Cheon and E. D. Cubuk, Scaling deep learning for materials discovery, *Nature*, 2023, **624**, 80-85.
25. B. Deng, P. Zhong, K. Jun, J. Riebesell, K. Han, C. J. Bartel and G. Ceder, CHGNet as a pretrained universal neural network potential for charge-informed atomistic modelling, *Nat. Mach. Intell.*, 2023, **5**, 1031-1041.
26. Z. Zou, Y. Zhang, L. Liang, M. Wei, J. Leng, J. Jiang, Y. Luo and W. Hu, A deep learning model for predicting selected organic molecular spectra, *Nat. Comput. Sci.*, 2023, **3**, 957-964.
27. X. Chen, S. Lu, Q. Chen, Q. Zhou and J. Wang, From bulk effective mass to 2D carrier mobility accurate prediction via adversarial transfer learning, *Nat. Commun.*, 2024, **15**, 5391.
28. Q. Zhu, Y. Huang, D. Zhou, L. Zhao, L. Guo, R. Yang, Z. Sun, M. Luo, F. Zhang and H. Xiao, Automated synthesis of oxygen-producing catalysts from Martian meteorites by a robotic AI chemist, *Nat. Synth.*, 2024, **3**, 319-328.
29. D. Zhang, P. Yi, X. Lai, L. Peng and H. Li, Active machine learning model for the dynamic simulation and growth mechanisms of carbon on metal surface, *Nat. Commun.*, 2024, **15**, 344.
30. C. Li, W. Yang, H. Liu, X. Liu, X. Xing, Z. Gao, S. Dong and H. Li, Picturing the Gap Between the Performance and US-DOE's Hydrogen Storage Target: A Data-Driven Model for MgH2 Dehydrogenation, *Angew. Chem.*, 2024, e202320151.
31. Y. Wu, C.-F. Wang, M.-G. Ju, Q. Jia, Q. Zhou, S. Lu, X. Gao, Y. Zhang and J. Wang, Universal machine learning aided synthesis approach of two-dimensional perovskites in a typical laboratory, *Nat. Commun.*, 2024, **15**, 138.
32. A. P. Shevchenko, M. I. Smolkov, J. Wang and V. A. Blatov, Mining Knowledge from Crystal Structures: Oxidation States of Oxygen-Coordinated Metal Atoms in Ionic and Coordination Compounds, *J. Chem. Inf. Mod.*, 2022, **62**, 2332-2340.
33. D. Cao, H. Xu, H. Li, C. Feng, J. Zeng and D. Cheng, Volcano-type relationship between oxidation states and catalytic activity of single-atom catalysts towards hydrogen evolution, *Nat. Commun.*, 2022, **13**, 5843.
34. M. J. Craig, F. Kleuker, M. Bajdich and M. García-Melchor, FEFOS: a method to derive oxide formation energies from oxidation states, *Catal. Sci. & Technol.*, 2023, **13**, 3427-3435.
35. X. Zhang, W. Meng, Y. Liu, X. Dai, G. Liu and L. Kou, Magnetic Electrides: High-Throughput Material Screening, Intriguing Properties, and Applications, *J. Am. Chem. Soc.*, 2023, **145**, 5523-5535.

36. J. A. Terrett, J. D. Cuthbertson, V. W. Shurtleff and D. W. MacMillan, Switching on elusive organometallic mechanisms with photoredox catalysis, *Nature*, 2015, **524**, 330-334.
37. J. H. Kroll, N. M. Donahue, J. L. Jimenez, S. H. Kessler, M. R. Canagaratna, K. R. Wilson, K. E. Altieri, L. R. Mazzoleni, A. S. Wozniak, H. Bluhm, E. R. Mysak, J. D. Smith, C. E. Kolb and D. R. Worsnop, Carbon oxidation state as a metric for describing the chemistry of atmospheric organic aerosol, *Nat. Chem.*, 2011, **3**, 133-139.
38. R. S. Mulliken, Electronic Population Analysis on LCAO–MO Molecular Wave Functions. I, *J. Chem. Phys.*, 1955, **23**, 1833-1840.
39. R. F. Bader, A quantum theory of molecular structure and its applications, *Chem. Rev.*, 1991, **91**, 893-928.
40. R. Bader, Atoms in Molecules: A Quantum Theory. USA: Oxford University, Press, *1994*, 1990, **32**.
41. J. Wu, D. Yu, S. Liu, C. Rong, A. Zhong, P. K. Chattaraj and S. Liu, Is it possible to determine oxidation states for atoms in molecules using density-based quantities? An information-theoretic approach and conceptual density functional theory study, *J. Phys. Chem. A*, 2019, **123**, 6751-6760.
42. J. E. Penner-Hahn, X-ray absorption spectroscopy in coordination chemistry, *Coord. Chem. Rev.*, 1999, **190**, 1101-1123.
43. J. Rehr and A. Ankudinov, Progress in the theory and interpretation of XANES, *Coord. Chem. Rev.*, 2005, **249**, 131-140.
44. J. Yano and V. K. Yachandra, X-ray absorption spectroscopy, *Photosynth. Res.*, 2009, **102**, 241-254.
45. P. W. Anderson, More Is Different: Broken symmetry and the nature of the hierarchical structure of science, *Science*, 1972, **177**, 393-396.
46. R. B. Laughlin and D. Pines, The theory of everything, *PNAS*, 2000, **97**, 28-31.
47. M. Jansen and U. Wedig, A piece of the picture--misunderstanding of chemical concepts, *Angew. Chem. Int. Ed. Engl.*, 2008, **47**, 10026-10029.
48. J. C. Golden, V. Ho and V. Lubchenko, The chemical bond as an emergent phenomenon, *J. Chem. Phys.*, 2017, **146**, 174502.
49. I. F. Leach and J. E. Klein, Oxidation States: Intrinsically Ambiguous?, *ACS Cent. Sci.*, 2024, **10**, 1406-1414.
50. M. Kaupp and H. G. von Schnering, Formal oxidation state versus partial charge—a comment, *Angew. Chem. Int. Ed. Engl.* 1995, **34**, 986-986.
51. I. D. Brown, Recent developments in the methods and applications of the bond valence model, *Chem. Rev.*, 2009, **109**, 6858-6919.
52. I. D. Brown, *The chemical bond in inorganic chemistry: the bond valence model*, Oxford university press, 2016.
53. N. Fu, J. Hu, Y. Feng, G. Morrison, H. C. z. Loye and J. Hu, Composition Based Oxidation State Prediction of Materials Using Deep Learning Language Models, *Adv. Sci.*, 2023, **10**, 2301011.
54. S. P. Ong, W. D. Richards, A. Jain, G. Hautier, M. Kocher, S. Cholia, D. Gunter, V. L. Chevrier, K. A. Persson and G. Ceder, Python Materials Genomics (pymatgen): A robust, open-source python library for materials analysis, *Comput. Mater. Sci.*, 2013,

**68**, 314-319.

55. T., Mueller, J. Montoya, W. Ye, X. Lei, L. Hung, J. Hummelshøj, M. Puzon, D. Martinez, C. Fajardo, R. Abela, An electrochemical series for materials, *PNAS.*, 2024, **121** (38), e2320134121.
56. A. Jain, S. P. Ong, G. Hautier, W. Chen, W. D. Richards, S. Dacek, S. Cholia, D. Gunter, D. Skinner and G. Ceder, Commentary: The Materials Project: A materials genome approach to accelerating materials innovation, *APL Mater.*, 2013, **1**.
57. S. Kirklin, J. E. Saal, B. Meredig, A. Thompson, J. W. Doak, M. Aykol, S. Rühl and C. Wolverton, The Open Quantum Materials Database (OQMD): assessing the accuracy of DFT formation energies, *npj Comput. Mater.*, 2015, **1**, 1-15.
58. M. Suvarna and J. Pérez-Ramírez, Embracing data science in catalysis research, *Nat. Catal.*, 2024, 1-12.
59. P. Pyykkö, Additive covalent radii for single-, double-, and triple-bonded molecules and tetrahedrally bonded crystals: a summary, *J. Phys. Chem. A*, 2015, **119**, 2326-2337.
60. L. Pauling, The principles determining the structure of complex ionic crystals, *J. Am. Chem. Soc.*, 1929, **51**, 1010-1026.
61. C. Tantardini and A. R. Oganov, Thermochemical electronegativities of the elements, *Nat. Commun.*, 2021, **12**, 2087.
62. S. Kearnes, K. McCloskey, M. Berndl, V. Pande and P. Riley, Molecular graph convolutions: moving beyond fingerprints, *J. Comput. Aided Mol. Des.*, 2016, **30**, 595-608.
63. W. Jin, C. Coley, R. Barzilay and T. Jaakkola, Predicting organic reaction outcomes with weisfeiler-lehman network, *NeurIPS*, 2017, **30**.
64. T. Xie and J. C. Grossman, Crystal graph convolutional neural networks for an accurate and interpretable prediction of material properties, *Phys. Rev. Lett.*, 2018, **120**, 145301.
65. C. W. Coley, W. Jin, L. Rogers, T. F. Jamison, T. S. Jaakkola, W. H. Green, R. Barzilay and K. F. Jensen, A graph-convolutional neural network model for the prediction of chemical reactivity, *Chem. Sci.*, 2019, **10**, 370-377.
66. S. Ryu, Y. Kwon and W. Y. Kim, A Bayesian graph convolutional network for reliable prediction of molecular properties with uncertainty quantification, *Chem. Sci.*, 2019, **10**, 8438-8446.
67. C. Kang, H. Zhang, Z. Liu, S. Huang and Y. Yin, LR-GNN: A graph neural network based on link representation for predicting molecular associations, *Brief. Bioinform.*, 2022, **23**, bbab513.
68. W. L. Hamilton, R. Ying and J. Leskovec, Representation learning on graphs: Methods and applications, *arXiv preprint arXiv:1709.05584*, 2017.
69. M. Henaff, J. Bruna and Y. LeCun, Deep convolutional networks on graph-structured data, *arXiv preprint arXiv:1506.05163*, 2015.
70. P. Velickovic, G. Cucurull, A. Casanova, A. Romero, P. Lio and Y. Bengio, Graph attention networks, *Stat*, 2017, **1050**, 10-48550.
71. Z. Xiong, D. Wang, X. Liu, F. Zhong, X. Wan, X. Li, Z. Li, X. Luo, K. Chen and H. Jiang, Pushing the boundaries of molecular representation for drug discovery with the graph attention mechanism, *J. Med. Chem.*, 2019, **63**, 8749-8760.
72. J. Gilmer, S. S. Schoenholz, P. F. Riley, O. Vinyals and G. E. Dahl, Neural Message

Passing for Quantum Chemistry, International conference on machine learning, *PMLR,* 2017, 1263-1272.
73. L. Breiman, Random forests, *Machine learning*, 2001, **45**, 5-32.
74. T. Chen and C. Guestrin, Xgboost: A Scalable Tree Boosting System, *Proceedings of the 22nd ACM SIGKDD,* 2016, 785-794.
75. C. Zhang, D. Song, C. Huang, A. Swami and N. V. Chawla, Heterogeneous Graph Neural Network, *Proceedings of the 25th ACM SIGKDD,* 2019, 793-803.
76. Z. Shui and G. Karypis, Heterogeneous Molecular Graph Neural Networks for Predicting Molecule Properties, 2020 *IEEE ICDM,* 2020, 492-500.

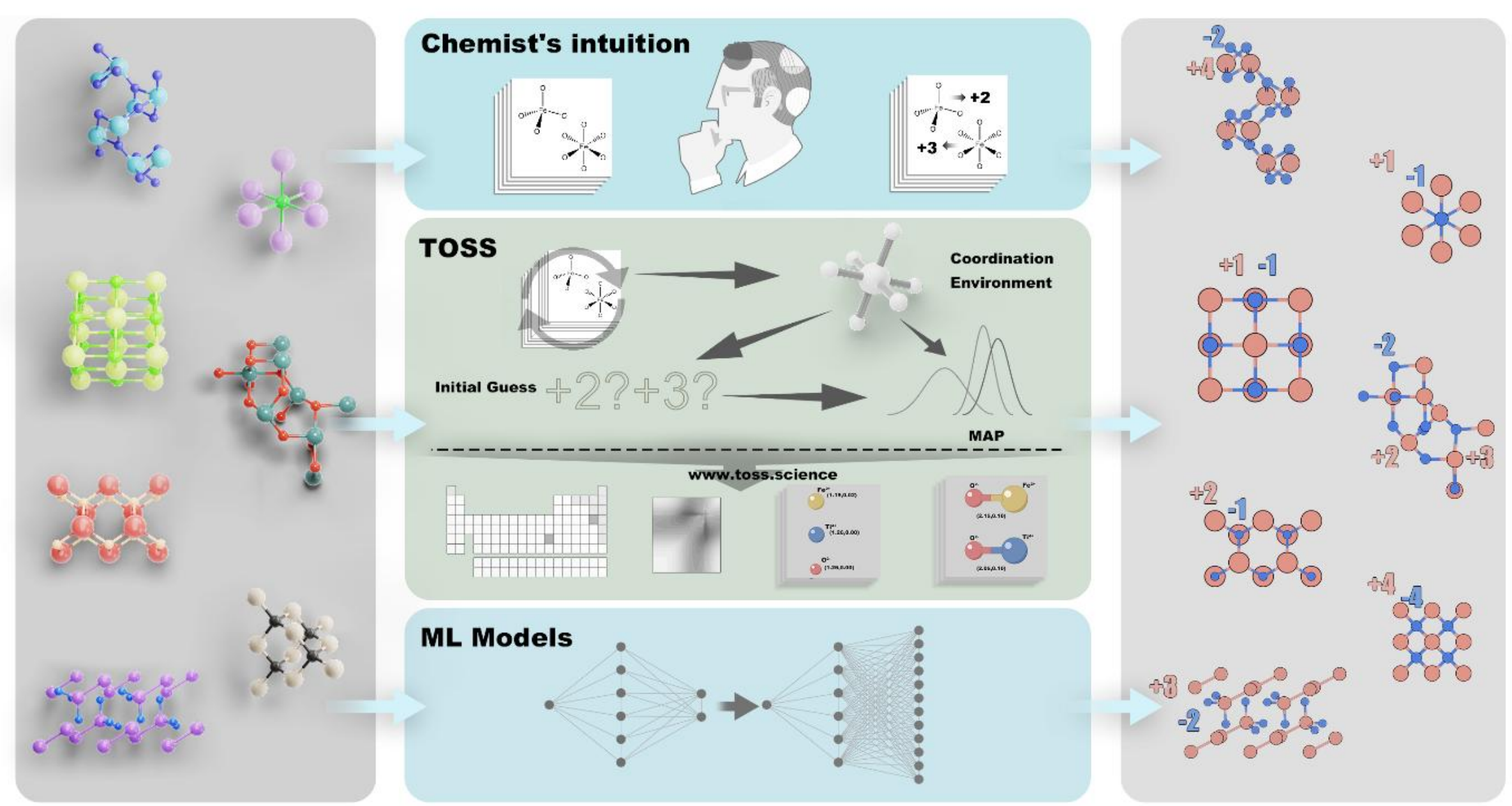

(TOC)